\documentstyle[natbib,color,amssymb,endnotes,doublespace,verbatim,footmisc,times,12pt]{article}

\newcommand{\revi}[1]{#1}
\newcommand{\secondrevi}[1]{#1}

\let\footnote=\endnote
\begin{document}
\title{A Puzzle about Further Facts\footnote{This paper will appear in {\em
    Erkenntnis}.}}
\author{Vincent Conitzer\\Duke University}
\date{}
\maketitle

\begin{abstract}
  In metaphysics, there are a number of distinct but related questions
  about the existence of ``further facts''---facts that are contingent
  relative to the physical structure of the universe.
  These include further facts about qualia, personal identity, and time.
  In this article I provide a sequence of examples involving computer
  simulations, ranging from one in which the protagonist can clearly
  conclude such further facts exist to one that describes our own
  condition.  This raises the question of where along the sequence (if at
  all) the protagonist stops being able to soundly conclude that
  further facts exist.
\end{abstract}

\noindent {\bf Case A.}  Fonda has just attended an inspiring department
colloquium. On her way out, she absentmindedly takes a wrong turn and
wanders into a computer lab.  She approaches one of the computers.  On it,
a simulation of a number of humanoid agents in a virtual universe is being
run.  The perspective of one of these agents---call it Alpha---is
continuously displayed on the monitor.  Fonda is enthralled and continues
to watch from Alpha's perspective.  Because Alpha takes a large variety of
interesting actions in the simulated universe, it does not take Fonda long
to learn the laws of physics governing it.  Moreover, Fonda---whose
undergraduate degree was in computer science---can easily imagine how one
would write the code for simulating the environment according to these
laws. Then, she has the following thought. There must be some {\em
  additional} code, beyond the code that executes the simulated universe's
laws of physics (and the code that gives its initial conditions).  Namely,
there must also be some code that governs the displaying of Alpha's
perspective on the screen.  This additional code could in principle be
changed without changing any of the code governing the physics and the
initial conditions.  For example, what is currently displayed as red on the
monitor could be displayed as blue instead, by changing only the code
governing the display.  As another example, this code could be changed to
display the perspective of a different agent instead. (It is possible that
the other monitors in the lab display the perspectives of the other
agents---she has not checked---but even so, one could change which agent's
perspective gets displayed on which screen.) Alternatively, the code could
be changed to display nothing at all (with the simulation nevertheless
running unchanged). Fonda thus concludes there are some ``further facts''
to this universe, in that there is additional code beyond that governing
the laws of physics and the initial conditions.  This additional code is,
in a sense, {\em contingent} relative to the other code.\footnote{Due to
  the existence of the additional code, these further facts are in fact
  {\em ontologically} further facts, as opposed to merely epistemologically
  further
  facts.  For more on the distinction, see, e.g.,~\citet{Chalmers10:Character}.}\\

The ``further facts'' that Fonda concludes are present in Case A are
related to those considered in the literature on metaphysics and philosophy
of mind.  The first example of a further fact in the case is related to
qualia and the possibility of inverted spectra.\footnote{See,
  e.g.,~\citet{Shoemaker83:Inverted} and~\citet{Block90:Inverted}.  The
  literature on inverted spectrum (and closely related) scenarios, their
  possibility, and their implications is, of course, vast.  To keep the
  length of this paper reasonable and avoid distraction from the main
  issues, I will omit a detailed review of, and comparisons to, the
  specific scenarios and arguments in this literature. I hope that the
  reader familiar with this literature can easily fill in the blanks.}  It
especially relates to strong versions of the inverted spectrum scenario
where qualia do not supervene on the physical, i.e., where two
microphysically identical twins nevertheless have inverted spectra.  Stated
otherwise, a variant of the question for which the case above is arguably
especially relevant is: did the laws of our universe (and its initial
conditions\footnote{For ease of exposition, I will assume that the laws of
  each universe are such that the initial conditions completely determine
  the physical structure of the universe. However, this is not essential to
  the arguments in this paper.}) {\em necessitate} that when I see red
things they phenomenally appear the way they do, or would they have allowed
for them to appear the way blue things do now?  Whether the phenomenal
nature of color perception is contingent (holding fixed the physical laws
of our universe and its initial conditions) has been the subject of much
philosophical debate.  Fonda's question is clearly related, though it would
seem that in her case she is entirely {\em right} to conclude that the way
things in the simulated universe appear to her is contingent, even holding
fixed the physical laws and initial conditions of the simulated universe.
The analogy between Fonda's questions in Case A and the standard ones from
the metaphysics and philosophy of mind literature breaks down at some
points, and we will explore this in what follows.  Before we do so, let us
consider the other example, Fonda's question of why it is Alpha, and not
some other agent, whose perspective is displayed.  This is related to
questions in metaphysics about personal identity and the self. Most
closely, it is related to the question of whether ``I could have been
someone else,'' and the closely related question of whether it is
contingent that {\em this} perspective is the present one.\footnote{Whether
  one believes that these two questions get at the same issue will depend
  on one's interpretation of them, perhaps especially of the former.  The
  former question has featured prominently in the literature on whether
  imagination provides a reliable guide to possibility.  Namely, if I can
  imagine myself being (say) Napoleon, then should we not conclude that
  {\em I could have been Napoleon}?  But it is difficult to make sense of
  this conclusion.  One way to do so is to interpret ``I'' as referring to
  a Cartesian self, and ``Napoleon'' as referring to an empirical self.
  See~\citet{Williams73:Imagination} for discussion of these points,
  or~\citet{Ninan16:Imagination} for a more recent discussion and further
  references.  Of course, most contemporary philosophers will eschew such
  an interpretation.  The latter question avoids Cartesian dualism by
  focusing on the {\em presence} of the experience, rather than on which
  entity has the experience.  This approach is closely related to the
  theory of ``egocentric presentism'' proposed
  by~\citet{Hare07:Self,Hare09:On} (see also~\citet{Hare10:Realism} and the
  closely related subjectivist theory laid out
  by~\citet{Merlo16:Subjectivism}), which is a subtle form of solipsism
  according to which only one human being's perspective is ``present.''
  Other recent work on these questions includes that
  by~\citet{Johnston11:Surviving} (e.g., the section ``Am I Now
  Contingently Johnston?'', pages 151-154) and~\citet{Hellie13:Against},
  who discusses the ``vertiginous question'' of why {\em Hellie} is the
  human being whose experiences appear ``live.''  \citet[page
  62]{Valberg07:Dream}, in support of similar ideas, discusses in detail
  the example of himself having a dream in which he occupies the
  perspective of someone other than himself, even though he---meaning,
  Valberg---is one of the characters in the dream.
  (\citet{Ninan16:Imagination} discusses similar ``Lakoff cases.'')}

Another type of further fact that could be included in the discussion is
that of further facts about time.  Fonda might ask herself why {\em this
  point in the simulation's timeline} is being displayed to her right now,
as opposed to another point in simulated time.  Also, she might ask herself
why the simulation runs at the {\em rate} that it does, as opposed to (say)
twice as fast.\footnote{Again, the literature addressing apparently related
  questions in the philosophy of time is vast, and I hope that the reader
  familiar with this literature can easily draw the connections.  For
  relatively recent references, see,
  e.g.,~\citet{Balashov05:Times,Zimmerman05:ATheory,Olson09:Rate},
  and~\citet{Skow11:On}.} Here, it may not be clear that things could have
been different.  For example, perhaps the simulation simply started running
when the program was first executed and it is running at the fastest
possible rate on the hardware provided.  On the other hand, the entire
simulation may have been precomputed from beginning to end, so that a type
of block universe is already stored in computer memory and Fonda is just
watching a replay of some part of it. In this case, there must indeed be
some further code governing which temporal part is replayed and at which
rate the replay runs.  While I believe that the cases presented here may
indeed provide some helpful insights for the metaphysics of time, things
are clearer for the other types of further facts.  Hence, I will avoid
discussion of time in what follows.

We are now ready to introduce the next case.\\

\noindent {\bf Case B.}  This case proceeds similarly to Case A, though
with an important difference.  Instead of plain monitors, the computer lab
now has sophisticated virtual reality (VR) headsets.  Fonda puts on one of
these that shows her the perspective of Alpha.  The VR system is so
remarkably good that Fonda spends a long time using it and becomes
completely engrossed---so engrossed that she completely forgets the world
outside the simulation, her own identity in it, and, we may suppose, even
basic facts such as what the color of grass is, or even the existence of
such a thing as grass at all (assuming there is no such thing in the
simulated universe). All that is left to her is the simulation, displayed
from Alpha's perspective---so that presumably she feels rather identified
with Alpha.  Again, she quickly learns the laws of physics in this
universe.  Then she has the following thought.  There must be further facts
to this universe, namely the ones concerning my own perspective in it.  The
(simulated) sky could have appeared to me in the color in which the
(simulated) ground appears to me now, without the fundamental laws of the
universe changing.  Moreover, there must be further facts regarding my
identity---why is it {\em this} perspective that appears (to
me\footnote{\label{fo:me}It should be pointed out that the referent of
  ``me'' is not clear here.  Since by assumption Fonda no longer remembers
  her life outside the simulation and she feels identified with Alpha, one
  could argue that for the thought to make sense from her perspective,
  ``me'' should refer to Alpha, in which case we end up with the familiar
  uninteresting question of why Alpha's perspective (rather than Beta's)
  should appear to Alpha.  On the other hand, she could take ``me'' to
  refer to some abstract observer, one that is difficult for her to
  identify, rather than a clearly identifiable agent in the (simulated)
  universe.  If she were to do so, it would make more sense for the
  referent to be Fonda, i.e., the human being wearing the VR headset, even
  though she is no longer aware of the existence of such a human being.
  And then, the question does have a nontrivial answer that involves the
  code governing the simulation and its display on the VR system. As yet
  another alternative, we can just leave out ``to me'' altogether---that is
  why this phrase is in parentheses.  Again, in this particular context,
  this seems to be a sensible question with a nontrivial answer. \revi{(See
    also Endnote~\ref{fo:concepts} on phenomenal concepts.)}})
and not that of some other agent?\\

Finally, Case C returns to day-to-day life.\\

\noindent {\bf Case C.}  In this version, Fonda does not walk into any
computer lab; she just walks outside and experiences the world as we
normally do.  She knows the laws of physics well from her undergraduate studies and nothing in the world seems mysterious to her (unresolved questions in physics aside).  Then, she has a thought just like the one in Case B, but now about our own familiar universe.  Why does the sky appear to me the way it does?  Why is it {\em this} perspective that appears (to me)?  There must be further facts beyond the laws of physics and any initial conditions.\\

In each case, Fonda reaches the conclusion that there are further facts to
the universe at hand.  In which of these cases is her conclusion justified?
It is worth emphasizing that the question is {\em not} whether there are
actually further facts, but rather whether the reasoning that leads her to
conclude this is sound.  In the same way as it is possible to give a wrong
proof for a (true) theorem, in principle Fonda's belief in further facts can
fail to be justified even if there are in fact further facts in these
cases.  Now, there are four possibilities:
\begin{enumerate}
\item Her reasoning is not sound in Case A.
\item Her reasoning is sound in Case A, but not in Case B.
\item Her reasoning is sound in Case B, but not in Case C.
\item Her reasoning is sound in Case C.
\end{enumerate}
It is straightforward to check that these four options are exhaustive in
the sense that at least one of them must hold.\footnote{Letting $+X$ denote
  that her reasoning is sound in case $X$ and $-X$ that it is not, there
  are $2^3=8$ combinations; $-$A$-$B$-$C,$-$A$-$B$+$C,$-$A$+$B$-$C, and
  $-$A$+$B$+$C are covered (at least) under 1, $+$A$-$B$-$C and
  $+$A$-$B$+$C under 2, $+$A$+$B$-$C under 3, and $+$A$+$B$+$C under 4.
  Alternatively, it is not hard to see that (for example) the negation of
  the first three possibilities implies the fourth.}  \revi{One may of
  course choose Option 4, having been convinced by the sequence of cases
  (or already believing prior to picking up this paper) that we are
  justified in concluding that there are further facts in our own world. I
  have little to say that is new about advantages and disadvantages of such
  a view, so the remainder of the paper is devoted to the following
  question.}  If we believe that \revi{Option} 4 is false, then which of
the first three \revi{options} is most plausible?  \revi{In what follows, I
  argue that Option 2 is the most appealing of the three, though attempts
  to decisively establish it as correct lead us to variants of known
  arguments about qualia and personal identity.  The exercise does cast a
  new light on these arguments, in particular clarifying some of their
  epistemological aspects.  It also demonstrates commonalities among
  various types of putative further facts that I believe have not been
  sufficiently appreciated in the literature.}

\section*{Option 1: Fonda's reasoning in Case A is not sound}

This, to me, seems the least appealing option of the three, so I will not
spend much space on it.  Fonda's reasoning seems entirely sound to me: if
the simulation is displaying on the monitor, there must in fact be some
code that governs this display.  It would certainly be possible to write
code for the simulation without any instructions to display anything on the
screen, but then the simulation would just run
silently\secondrevi{\footnote{\secondrevi{By using the word ``silently'' I
      do not intend to take any stance on whether and in what sense there
      might be such a thing as Alpha's {\em own} experience; this is
      irrelevant to the arguments presented here, which concern Fonda's
      experience.  See also Endnote~\ref{fo:fondanotalpha}.}}} on the
machine without any output.  This is in fact a mistake programmers make on
occasion: they write the code (for, say, computing $2^n$ as a function of
$n$) correctly but forget to write instructions to display the result to
(say) the screen. Such a mistake is typically easily corrected by adding a
line to the code.\revi{\footnote{\revi{It is not required here that the
      code for running the simulation and the code governing the display
      are neatly separated. Even if they are intermingled in horribly messy
      ways, {\em somewhere} in the code there must be commands of roughly
      the
      following form:\\
      \noindent \texttt{display(x,y,z)}\\
      \noindent indicating that on the screen the pixel at coordinates $x$
      and $y$ is to be given color $z$.  Removing all (and only) these
      commands will result in nothing being displayed, even though the
      simulation of the physics is running.  Alternatively, replacing $z$
      by $z+1$ in every such command will result in all the displayed
      colors changing a bit.

      Is it conceivable that the code was written (and, perhaps, for some
      reason {\em had} to be written) in a strange programming language
      that would prevent changing the colors?  Or that the monitor for some
      reason (say) cannot display large amounts of red at the same time,
      necessitating red to be used for a particular simulated wavelength?
      Perhaps, but it is easy to argue that Fonda has strong reason to
      believe that a sensible programming language was used and that the
      monitor does not have strange constraints.  In any case, here, in
      Case A, we can simply sidestep these concerns by specifying that
      Fonda knows the programming language and the type of monitor used.}}}

One could argue for Option 1 by arguing that in fact, there are no further
facts in Case A.  Such an argument might proceed as follows.  
\begin{quote}
  We should distinguish between two claims.  One is that the qualia
  associated with seeing an object supervene (only) on facts about the
  physical properties of that object. Let us call these the ``narrow''
  physical facts.  The other is that they supervene on facts about the
  physical properties of the object being viewed, those of the observer
  viewing the object, and those of anything else mediating the viewing.
  Let us call these, collectively, the ``broad'' physical facts.  The
  former claim is untenable, for example because the qualia are different
  when the observer is color blind or the air between the object and the
  observer is hazy.  It is the latter claim that is of interest.  And in
  Case A, the broad physical facts include facts about the code governing
  the display, the monitor itself, Fonda's eyes and brain, etc.  Hence,
  there is no reason to think that there are any further facts in Case A.
\end{quote}
However, this argument relies on misunderstanding the sense in which the
phrase ``further fact'' is being used here, which is quite modest.  The
point is that Fonda's experience, or even just what appears on the monitor,
is not fully determined by the physical---{\em in the sense of the
  simulated physics}---facts of the simulated universe.  These include
facts about the simulated objects, Alpha, and anything mediating the
viewing {\em within} the simulated universe. They do {\em not} include
physical (in the common sense) facts about the monitor, Fonda's body, and
the space between them.  They also do not include facts about the
additional code governing the display.  Again, the relevant physics is the
physics of the simulation, not the physics of the broader
world.\secondrevi{\footnote{\label{fo:fondanotalpha}\secondrevi{But then,
      could it similarly be the case that the relevant {\em experience} is
      Alpha's, not Fonda's?  One can argue that Alpha's experience is fully
      determined by the simulated physics, so that no further facts are
      needed to explain Alpha's experience.  However, this will not resolve
      the puzzle considered in this paper.  It is not clear under what
      conditions there is something it is like to be Alpha---that is, not
      to be someone to whom Alpha's perspective is being displayed, but to
      really {\em be} Alpha---but in any case this is irrelevant to the
      issues we investigate here.  In Cases A and B as I have specified
      them, the conclusions about further facts are reached by Fonda, not
      Alpha, on the basis of Fonda's experience, not Alpha's.  So Fonda's
      experience is the relevant one.  But, one might ask, is Fonda in Case
      C perhaps more similar to Alpha in Case B than to Fonda in Case B?
      At least for some aspects of these three entities, this is surely
      true.  Does this mean that there is a gap between Cases B and C, and
      that to close the gap we should {\em modify} Case B to have Alpha,
      not Fonda, conclude that there are further facts?  No.  For our
      purposes, it is not important that Fonda in Case B and Fonda in Case
      C are similar in every aspect.  All that matters is that their {\em
        epistemic} situations are similar across these cases.  I will
      discuss this in more detail in the section on Option 3.  See also
      Endnote~\ref{fo:me} on what the referent of ``me'' is.}}}  It is
clear that there are further facts in Case A in the modest sense of being
contingent relative to just the facts about the simulated physics, and this
modest sense is the one of interest in this paper.  Why this interpretation
is the one of interest is made clear by considering the analogous move of
using an immodest interpretation in Case C.  This move would result in
arguments such as the following.
\begin{quote}
  Even if experiences were had by souls outside the world through some
  process mediating between brains in the world and souls outside it, then
  we should simply consider a broader physics that includes the souls and
  the mediating process. By doing so the experiences again supervene on the
  broader physical facts, so there is still no evidence for further facts.
\end{quote}
Clearly this argument is unsatisfactory; in arguing against further facts, we mean
to argue against the existence of things such as extraworldly souls, not
to accommodate them through a technical maneuver.  While all this may seem
rather obvious, it is important to keep straight, especially in Case B,
where the relevant physics is still the physics of the simulation---which,
in that case, is the only physics of which Fonda is aware.

I will now skip to Option 3 before returning to Option 2.

\section*{Option 3: Fonda's reasoning is sound in Case B but not in Case C}

While this option seems more appealing to me than Option 1, it still seems
difficult to argue for it.  Key to this difficulty, of course, is that by
assumption, Fonda has forgotten {\em everything} about the outside world in
Case B.  If she retains some memory of the outside world, the case will
reduce to one that is not substantively different from Case A.

Is Case B substantively different from Case C?  Of course: in Case B the
universe under consideration is a simulation in a larger universe.  But are
the two cases substantively different in terms of Fonda's epistemic
situation?  This is what seems difficult to argue.  Whether we have reason
to believe that we are not a brain in a vat or (in) a computer simulation
is a topic that has been explored at length in the literature.
\citet{Bostrom03:Are} has argued that there is a large probability that we
are in fact in a computer simulation, under some assumptions including that
posthuman civilizations are likely to be reached and likely to run a large
number of such simulations.  In contrast,~\citet{Markosian14:Do} has argued
that all the evidence speaks in favor of the external world being real;
evidence in favor of being a brain in a vat would be exemplified by a major
glitch in the simulated environment.  For our purposes, it is not necessary
to resolve this debate. What matters is not whether Fonda is justified in
believing that the world around her is real (in the sense of not being a
simulation) in either Case B or C, but rather whether there is a
substantive difference between these two cases in terms of her epistemic
situation.

Let us simply assume that there are no glitches in Case B.  Furthermore, at
least in principle, the simulation in Case B could provide Fonda with a very
rich experience.  As Case B has been described so far, Fonda is not able to
take {\em actions} in it; she is just observing.  This indeed constitutes a
difference between what Fonda observes in Cases B and C. It is not
immediately clear to me whether and how this particular difference is
relevant to the soundness of her argument for further facts.  In any case,
we can easily modify the example to give her some control over Alpha's
actions (with perhaps other human beings who similarly wandered into labs
controlling the other agents in the simulation).

Perhaps more interestingly, she may wonder about the place of her own
thoughts in the simulated universe.  In our own world, we have reasons to
believe that our thoughts are generated by our brains.  If a similar
account does not seem reasonable in the simulated universe---for example,
because there does not seem to be any physical structure in it capable of
generating these thoughts---she may conclude that such a structure must
exist somewhere outside of her observable universe, and from there it is a
short step to conclude the existence of further facts.  Then again, we
could modify the example so that there appear to be brains inside the
simulated agents; we could even go so far as to imagine that, unbeknownst
to Fonda, her brain is being scanned while she is standing in the lab, and
what goes on in it is then reflected in Alpha's simulated
brain.\footnote{Neuroscience aside, in our own world we also observe, to
  some extent, how children learn to think.  If there is nothing analogous
  in the simulation, again this may raise suspicions about there being an
  ``outside'' world where the ability to think rationally is acquired.
  Again, though, it does not seem difficult to modify the case
  appropriately, for example with Fonda and perhaps others having been in
  the virtual reality system since childhood.}  (A similar idea is
described by~\citet{Chalmers05:Matrix}.)

Overall, it seems difficult to draw a sharp distinction between Fonda's
epistemic situation in Cases B and C that cannot be addressed with a simple
modification of the cases.\revi{\footnote{\revi{One may ask whether this
      presupposes an internalist view of epistemic justification.  Might an
      externalist not argue that Fonda's belief in further facts is
      justified in Case B but not in Case C, because whether her belief is
      justified hinges on aspects of the external world?  For one, it may
      be that in Case B, Fonda's prior experiences outside the simulation
      contributed causally to her current thoughts about further facts,
      even though she is not currently aware of this.  (Examples in which
      one is not aware of exactly how one has come to believe something are
      common in the literature about internalist vs.~externalist views of
      epistemic justification; see, e.g.,~\citet{Goldman09:Internalism}.)
      If so, externalists, and even some internalists, may consider Fonda's
      belief in further facts justified in Case B but not in Case C.
      However, we can simply specify that no such causal link exists in
      Case B---say, Fonda's brain has rewired itself from scratch after
      entering the simulation.  Given this additional detail, it seems few
      externalists would hold that Fonda's belief in further facts is
      justified in Case B but not in Case C. For example, if we consider
      reliable process theory (see, e.g.,~\citet{Goldman79:What} for a
      classic version), it is not clear in what sense the process leading
      to Fonda's belief should be more reliable in Case B than in Case
      C.}}} Of course, this is precisely the point of Case B, to make
Fonda's epistemic situation in it essentially identical to that in Case C;
and if we can in fact succeed at this, then Option 3 fails.  This leaves us
with Option 2.

\section*{Option 2: Fonda's reasoning is sound in Case A but not in Case B}

This appears to me the most attractive of the three options.  {\revi A
  first attempt at an} argument proceeds as follows.  In Case A, Fonda
recognizes (say) that the color of the sky in the simulation is the same as
the color of grass in the outside world (i.e., green), whereas it could
just as well have been displayed as the color of the sky in the outside
world (i.e., blue).  That is, the {\em correspondence} between colors in
the simulation (as displayed on the screen) and colors in the outside world
clearly could have been different, without the code that governs the laws
of physics and the initial conditions in the simulation being any
different.  Thus there is clearly a ``further fact'' present, consisting in
the additional code that governs how a perspective is displayed on the
monitor.  In Case B, however, Fonda cannot recognize the existence of any
such correspondence, because she no longer remembers the outside world.
Hence---so the argument goes---she cannot conclude further facts exist.

Now, this argument does not seem entirely satisfactory to me.  Even in Case
B, it seems entirely possible for Fonda to imagine a scenario where the
color of the sky (in the simulation, though she does not know it is a
simulation) would be the color that the grass is now (in the simulation),
and vice versa.  This is the familiar inverted spectrum scenario, except in
this case, by virtue of all this taking place in a virtual reality system,
it is clearly {\em true} that the spectrum could be inverted; all this
would require is some changes to the code governing the display.
Nevertheless, it is still possible that Fonda's belief that the spectrum
could have been inverted is not {\em justified}.  Having forgotten about
the outside world, she certainly does not know about the simple
mechanism---changing a few lines of code in the outside world---by which
the spectrum could indeed be inverted.  But \revi{few would argue} that
awareness of a specific mechanism by which the spectrum could be inverted
is necessary to justify belief in the possibility of an inverted spectrum
\revi{(though it is clearly sufficient).

  Nevertheless, it seems that the physicalist, arguing that Fonda's belief
  is not justified, has arguments available in Case B that are unavailable
  in Case A.  The physicalist can argue that Fonda cannot be sure that
  experiential properties corresponding to her seeing the simulated sky are
  not, at bottom, physical properties.  (Again, here, ``physical'' refers
  to the physics of the environment, which we happen to know is simulated
  but she does not.)  Even in our own case (Case C), fleshing out such an
  argument and addressing immediate counterarguments requires substantial
  work; see, for example,~\citet{Hawthorne02:Advice}.  But I do not see
  that any additional obstacles to such an argument are introduced when
  moving from Case C to Case B.  In contrast, in Case A such an argument
  becomes untenable.  In that case, Fonda clearly knows that the
  experiential properties corresponding to her seeing the simulated sky are
  not, at bottom, properties of the simulated physics; she knows that the
  code governing the display, the monitor itself, her eyes and brain, etc.,
  are also involved.}

So, perhaps this all reduces to standard arguments about inverted spectra.
Perhaps Fonda cannot reasonably reject the possibility that the way colors
appear to her necessarily {\em emerges} from the laws of her universe.  If
so, it at least suggests that the debate on inverted spectra has been on
the right track.  But it also provides a clearer lens on these
arguments.\revi{\footnote{\label{fo:concepts}\revi{Of course, I do not
      claim that these examples have significant implications for every
      argument in the literature.  An exhaustive analysis of where they can
      provide insight is far beyond the scope of this paper, but, for
      instance, it is instructive to reconsider phenomenal concepts (for a
      survey article, see~\citet{Balog09:Phenomenal}), and especially their
      role in Chalmers' work (see,
      e.g.,~\citet{Chalmers96:Conscious,Chalmers03:Content}), in light of
      them.  Even for this, a thorough analysis is beyond the scope of this
      paper, but here is a sketch of how part of such an analysis might
      proceed.

      Phenomenal concepts are taken to pick out phenomenal qualities.
      However, in the context of examples with simulations (Cases A and B),
      we can define analogous concepts that simply pick out the color
      displayed on the screen (as opposed to the phenomenal color quality
      experienced by Fonda).  Even in Case B Fonda could herself form such
      a concept by means of imagination (``the color displayed on the
      screen assuming I am in a simulation'').  Standard arguments,
      including ones about the possibility of inverted spectra or zombies,
      can then be applied to these concepts instead of the phenomenal ones,
      and may become less controversial (since, e.g., we know we can invert
      the colors on a screen).  Then again, they may have less bite.  If
      all that such an argument allows Fonda to conclude in Case B is that
      an inverted spectrum---in the limited sense of the colors on a screen
      being inverted---is possible {\em if} she is in a simulation, then it
      is not clear how this by itself could justify an unconditional belief
      in the possibility of an inverted spectrum in the original phenomenal
      sense.  Fundamentally, the challenge for this approach seems to be
      that the new concepts are not infallible in the way that phenomenal
      concepts are widely held to be, and, relatedly, that they do not
      refer to something that can justify beliefs through acquaintance in
      the way that phenomenal qualities are widely held to be able to.
      (See also Endnote~\ref{fo:me} on what ``me'' could refer to for Fonda
      in Case B.)}}}  This is because unlike in the standard inverted
spectrum scenario, in this case it is clearly {\em true} that the spectrum
could have been inverted.  This, I believe, reduces the intuitive appeal of
the argument that which quale appears must supervene upon properties of the
physical world (and that therefore a strong type of inverted spectrum is
not possible).  It makes it clear that if this argument is to succeed, it
should be fundamentally epistemological in nature: it should argue just
that we cannot {\em know} that there is no such supervenience.  \revi{At
  least, this is so if our epistemic situation is sufficiently like that of
  Fonda in Case B, and it appears that it is, as discussed in the previous
  section.}

It is useful to note that even slightly nudging Case B towards Case A---for
example, allowing Fonda to remember only {\em that} she is in a simulation,
but effectively nothing else about the outside world, including even
whether her color experiences there were anything like the ones she is
experiencing now---would again allow her to soundly conclude that an
inverted spectrum in her simulated world is a genuine possibility.  This is
why it is important to be strict about Fonda not remembering {\em anything}
in Case B.

Next, let us consider further facts about personal identity and the self.
Again, in Case A, Fonda can soundly conclude that there are further facts
about how the perspectives of agents in the simulation are assigned to
monitors in the outside world.  Even if every agent's perspective is
displayed on some monitor, clearly the {\em correspondence}---on which
particular monitor each agent's perspective is displayed---could have been
different, without the code that governs the laws of physics and the
initial conditions of the simulation being any different.  Thus there is
clearly a further fact present, consisting in the additional code that
governs on which (if any) monitors each perspective is displayed.  Again,
however, in Case B, Fonda cannot recognize the existence of any such
correspondence, because she no longer remembers the outside world.  So one
might argue that in Case B she is not justified in believing in the
existence of further facts about personal identity and the self.

How satisfying is this argument?  Could she nevertheless, in Case B,
imagine the perspective of an agent other than Alpha appearing to her?
Certainly it is {\em true} that a different agent's perspective could be
made to appear to her; all this would require is a change to the code
governing which perspective is displayed by the VR system.  But would she
be {\em justified} in believing that a different perspective could have
appeared to her?  Again, it seems that the moment we allow her to remember
even just the mere fact that she is in a simulation, even if she remembers
nothing else about her identity in the outside world, she can indeed
conclude that a different agent's perspective could have been made to
appear to her.  But we explicitly rule out such a memory in Case B.  Hence,
it is not clear that the notion of a different perspective appearing {\em
  to her} makes sense from Fonda's perspective.  For all she knows, she
{\em is} Alpha, and how could any perspective other than Alpha's appear to
Alpha?

It is interesting to note that this argument is not entirely analogous to
the corresponding argument regarding color appearance given earlier. It may
be plausible to Fonda that the way colors appear necessarily emerges from
the laws of the universe in which she finds herself.  But it seems
implausible that somehow Alpha's perspective, to the exclusion of any
other, necessarily emerges as the ``present'' one from these laws, given
that Alpha is just one agent among many similar ones as far as these laws
go.\revi{\footnote{\revi{To flesh this out, we may specify that all the
      agents in the simulation---Alpha, Beta, Gamma, \ldots---are
      objectively (from a standpoint within the simulation) extremely
      similar.  Then, given that the (simulated) physical laws treat
      similar agents similarly, these laws could not by themselves
      determine which agent's perspective becomes the present (displayed)
      one.  In contrast, the phenomenal properties of seeing blue are
      inherently different from those of seeing red, so it does not seem
      possible to make a similar argument for the case of such qualia.
      Conceivably to Fonda in Case B, the phenomenal properties of seeing
      blue are somehow inherently linked to the physical properties of
      certain wavelengths in her environment.}}}  Instead, the argument
here relies on the possibility of her complete identification with Alpha.

Again, perhaps this all reduces to standard arguments about personal
identity and the self.  If so, then again, this on the one hand suggests
that the debate has been on the right track, while on the other hand also
casting a clearer lens on it.  This is because unlike in standard scenarios
in the literature on personal identity and the self, here it is clearly
{\em true} that Fonda ``could have been someone else''---i.e., she could
have had a different agent's perspective in the simulation displayed to her
on the VR system.  This highlights, again, that the problem fundamentally
has an important epistemological component.

\section*{Conclusion}

How can we avoid concluding that further facts exist in Case C, which
corresponds to our own world?  It seems that the approach most likely to
succeed is to argue that, while in Case A the conclusion of further facts
is justified, it is not in Case B.  Moreover, the most natural way to do so
is to counter the argument in favor of further facts in Case B in a way
that is similar to how arguments in favor of further facts in our own world
(Case C) are often countered.  But it appears that these counterarguments
lose at least some of their immediate intuitive appeal when moving from
Case C to Case B.  This is because, by construction, there are in fact
further facts in Case B, making it difficult to point out where exactly the
argument that further facts exist goes wrong.  Moreover, this argument is
essentially the same as in Case A, where presumably we do believe the
argument is correct.  Therefore, the counterargument needs to rely entirely
on Fonda's epistemic limitations in Case B.  Again, I believe that this is
the most natural way to avoid the conclusion that further facts exist.  But
I also believe that the counterargument is in need of further fleshing out.
\revi{In any case, if one agrees that Fonda's situation in Case B is
  epistemically sufficiently like ours (Case C), so that any arguments
  available to us against further facts should be available to Fonda in
  Case B as well, we obtain a nontrivial conclusion.  This is that we
  cannot {\em know} with certainty that qualia supervene on the physical
  facts, because after all, (say) an inverted spectrum is genuinely
  possible in Case B.\footnote{\revi{\label{fo:conceivability}Unlike for
      standard conceivability arguments (see,
      e.g.,~\citet{Chalmers10:Two}), where a key issue is whether the jump
      from conceivability to possibility can be made, it does not seem that
      possibility is at issue for Cases A and B.  Our current state of
      technology already enables at least Case A.  As for Case B, I am not
      aware of anyone having ever become so completely lost in a VR
      simulation, presumably at least in part due to remaining limitations
      of the technology.  But this technology is advancing rapidly with no
      apparent fundamental obstacles in its path.  Hence, it is hard to see
      what could keep Case B from being possible even in the near future.
      Moreover, inverting the spectrum or changing the identity of the
      displayed agent is clearly possible in Cases A and B. Unless an
      epistemic line can be drawn between Cases B and C, this seems to
      imply a strong type of epistemic possibility in our own case as
      well.}} At most, one can argue that we have no good reason to believe
  that they {\em fail} to supervene on the physical facts, and hence are
  not justified in concluding that there are further facts.}
 
One may take other routes.  I do not see how one could reasonably hold that
Fonda's argument in Case A is flawed.  On the other hand, perhaps one could
successfully argue that there is a relevant difference in Fonda's epistemic
situation between Cases B and C.  Of course, Case B is intended to be
epistemically as similar to Case C as possible, and for any remaining
potentially relevant epistemic difference between the two cases, it seems
we can modify Case B appropriately to make the difference go away.  One
might perhaps argue that Case B describes a scenario that is inherently
{\em impossible}---or at least that it would become so after sufficiently
many of these modifications.  But I see no convincing reason to think
so.\revi{\footnote{\revi{See also Endnote~\ref{fo:conceivability}.}}}

Debates about further facts are ancient and clearly I have not settled
them.  I do believe that the three cases presented here, besides putting a
modern spin on these questions, help to disentangle some of the different
aspects relevant to these debates.  They also allow us to treat different
types of further facts in a more uniform manner.

\revi{
\section*{Acknowledgments}

I am thankful to anonymous referees who provided especially thorough and
helpful comments, which significantly improved the paper.}

     \begingroup
     \parindent 0pt
     \parskip 1ex
     \theendnotes
     \endgroup

\end{document}